\documentclass[preprint,12pt]{elsarticle}
\usepackage{amssymb}
\usepackage{epsfig}
\journal{Physica A}
\begin{document}
\begin{frontmatter}
\title{Structural and topological phase transitions \\ on the German Stock Exchange}
\author[asien]{M. Wili{\'n}ski}
\ead{mateusz.wilinski@fuw.edu.pl}
\author[asien]{A. Sienkiewicz}
\ead{asien@okwf.edu.pl}
\author[asien]{T. Gubiec}
\ead{Tomasz.Gubiec@fuw.edu.pl}
\author[asien]{R. Kutner}
\ead{(for correspondence): erka@fuw.edu.pl}
\address[asien]{Institute of Experimental Physics Faculty of Physics, University of Warsaw  \\ Ho\.za 69, PL-00681 Warsaw, 
Poland}
\author[struz]{Z.R. Struzik}
\ead{zbigniew.struzik@p.u-tokyo.ac.jp}
\address[struz]{The University of Tokyo, Bunkyo-ku, Tokyo 113-8655, Japan}

\begin{abstract}
We find numerical and empirical evidence for dynamical, structural and topological phase transitions on the (German) 
Frankfurt Stock Exchange (FSE) in the temporal vicinity of the worldwide financial crash. Using the Minimal Spanning 
Tree (MST) technique, a particularly useful canonical tool of the graph theory, two transitions of the topology of 
a complex network representing FSE were found. First transition is from a hierarchical scale-free MST representing the stock 
market before the recent worldwide financial crash, to a superstar-like MST decorated by a scale-free hierarchy 
of trees representing the market's state for the period containing the crash. Subsequently, a transition is observed 
from this transient, (meta)stable state of the crash, to a hierarchical scale-free MST decorated by several star-like 
trees after the worldwide financial crash. The phase transitions observed are analogous to the ones we obtained earlier 
for the Warsaw Stock Exchange and more pronounced than those found by Onnela-Chakraborti-Kaski-Kert\'esz  
for S\&P 500 index in the vicinity of Black Monday (October 19, 1987) and also in the vicinity of January 1, 
1998. Our results provide an empirical foundation for the future theory of dynamical, structural and topological phase 
transitions on financial markets.
\end{abstract}
\end{frontmatter}

\section{Introduction}\label{section:hypothesis}

It is only since two decades, that the physicists have begun intensively to study structural and topological properties 
of complex networks \cite{DGM,KwDr} (and refs. therein) in order better to understand the mechanisms responsible for 
the evolution of complex systems. 

The theory of networks, or graphs, is based on the notion of vertices (or nodes) -- which can be identified with 
individual elements of the system, and edges (or links) -- which represent any connection between a pair of nodes.
Physicists have discovered that in most of real-life graphs, small and finite loops are rare and 
insignificant\footnote{Indeed, even if significant, renormalization of the graphs will collapse such loop structures, 
e.g., replacing the node triangles by effective vertices. Nevertheless, the stability and robustness of fully detailed real-life graphs 
should also be systematically studied.}. 
It can be conveniently assumed, that the topologies of many real world systems' graphs are locally dominated by trees 
of which the properties have been extensively exploited. 
Therefore, we decided on the Minimal Spanning Tree (MST) technique. It is a particularly useful, canonical tool in graph 
theory \cite{BeBo}, being a correlation based network without any loop \cite{RNM,BCLMVM,MS,BLM,VBT,KKK,TMAM,TCLMM}. 
For such a network the inter-node distance equals $d(i,j)=\sqrt{2(1-C(i,j))}$ for any pair of vertices. The 
transformation from the Pearson's correlation coefficient, $C(i,j)$, to distance, $d(i,j)$, is necessary because, the 
correlation coefficient does not obey the axioms of a metric (or even axioms of a subdominant ultrametric distance \cite{RNM}).

Prior to crash, the positive feedback or the herding effect dominates the behaviour of the stock market \cite{DiSo}. 
Hence, we can suppose that several stocks can be indirectly correlated, mainly over the trend forming a stock market bubble.
Such a trend could potentially be the main source of spurious correlations and (non-partial) Granger causality (usually difficult to calculate directly, cf. \cite{SMYHDFM} and refs. therein). We, therefore, substract the trend from all the assets' time series, following a well known approach \cite{MS}. 

In this work, we consider by numerical means, the evolution of $N=562$ companies quoted on the Frankfurt Stock Exchange\footnote{Notably, 
DAX contains only 30 largest companies.} (FSE), during the most intriguing period ranging from 2005-01-03 to 2008-08-12.
We divided this period into three sub-periods: (i) the first one ranging from 2005-01-03 to 2006-03-09 and consisting 
of 309 trading days, (ii) the next sub-period from 2006-04-20 to 2007-10-11 and consisting of 400 trading days, and 
(iii) the third one ranging from 2007-06-01 to 2008-08-12 and consisting of 313 trading days. This division, 
equipped with a little overlap between the second and third sub-periods, results in the most distinct MSTs. 
The differences rendered between subsequent states of the market obtained in this way, are most pronounced, as described in the following.

From the above mentioned 562 companies, only $N=466$ companies survived until the end of the first sub-period. At the 
end of the second and the third period there are $N=479$ companies, although several of them (more than $13$) are different companies. 
Indeed, during the market evolution from one sub-period to another, some vertices and edges may disappear in the 
corresponding network, while new vertices are created. Furthermore, the distances between vertices may also vary in 
time. Therefore, we regard the number of vertices and edges of the network as non-conserved quantities varying over 
time. Consequently, the characteristics of network's topology considered in this work as the most appropriate ones are time varying. 
These entail e.g. the mean occupation layer -- this quantity is more sensitive than, for instance, mean 
tree length\footnote{The mean tree length is a quantity which is quite often degenerated in respect to the complex network 
topology. That is, several different structures result in the same (or almost the same) mean length. This feature makes 
the mean tree length an inadequate tool for distinguishing between less and more centralised networks.} \cite{RNM,BLM,BR,TSC}.
This is because the logarithm of the mean occupation layer resembles nonequilibrium entropy of a topological complex network 
(see Section \ref{section:results}), which helps to identify the temporal key vertices in a complex network \cite{JulReh}.

We applied the MST technique to investigate transient behaviour of a low-order complex network during its evolution from 
a scale-free\footnote{In this work terms \emph{scale-free} and \emph{power law} we consider as synonyms.} topology representing 
the stock market hierarchical structure before the recent worldwide financial crash 
\cite{DiSo}, to a superstar-like tree (or superhub) decorated by scale-free hierarchy of trees (or hubs) -- representing 
a high-order market structure during the period containing the crash. Such a superstar-like MST corresponds with a relatively unstable (or 
metastable) state of financial market. This is because the structure of this state is much too ordered, compared with the equilibrium state. Subsequently, we found a transition from this unstable (or metastable) state to scale-free topology, decorated by a hierarchy of 
local star-like trees or hubs, again representing lower order market structure and topology directly after the worldwide financial crash. 

Similar types of transitions were found not only on the FSE but also on the Warsaw Stock Exchange (WSE) -- 
a complex network of 274 companies, quoted on the WSE throughout the period in question. Here, we omitted the results 
obtained for the WSE because they resemble those found for the FSE and they have been presented in Ref. \cite{SGKS}.
Both our results, that is for WSE and FSE, are much more pronounced than those found by Onnela, Chakraborti, Kaski, and Kert\'esz  
for 116 stocks of S\&P 500 index in the vicinity of Black Monday (October 19, 1987) \cite{OCKK1} and also in the vicinity of 
January 1, 1998 \cite{OCKK2}.

We foresee that our results, complementary in nature to the work by our colleagues reported in Ref. \cite{KwDr} (and 
refs. therein), may in the future, serve as a phenomenological foundation for modelling dynamic structural and 
topological phase transitions and critical phenomena (e.g. self-organised criticality \cite{PBX}) on financial markets 
\cite{DGM,DiSo,WH}.

\section{Results and discussion}\label{section:results}

A graph -- or a complex network -- representing the FSE was calculated for $N=466$ companies present on the FSE for 
the sub-period of time from 2005-01-03 to 2006-03-09 (covering 309 trading days), when the worldwide financial crash 
had not yet occurred \cite{DiSo}. For the construction of the MST, we here used Prim's algorithm \cite{PA}, which is 
quicker than Kruskal's \cite{PA,JK}, particularly for $N\gg 1$ which is the case here. However, both algorithms 
(and their various modified versions) are quite often used in this context.
\begin{figure}
\begin{center}
\bigskip
\includegraphics[width=120mm,angle=0,clip]{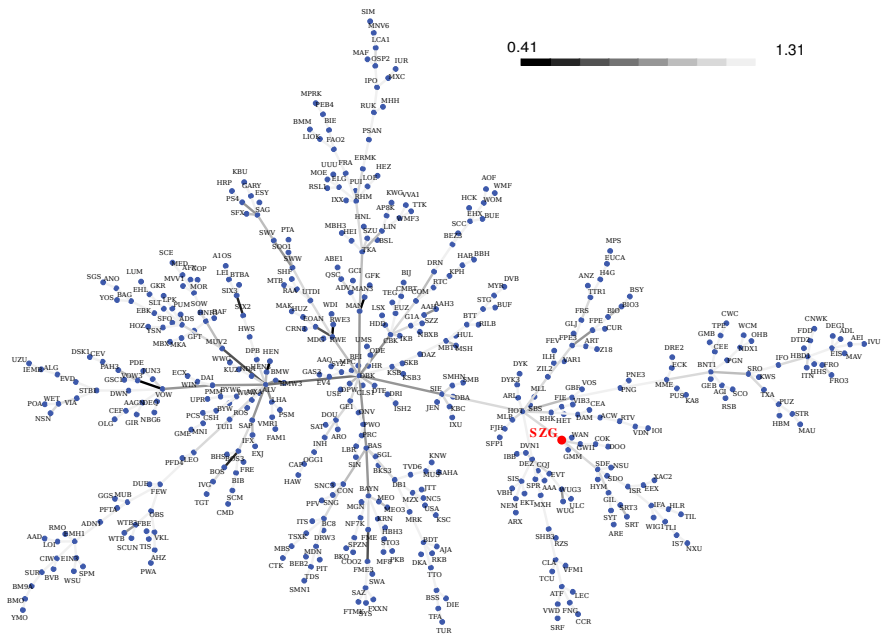}
\caption{The hierarchical MST associated with the FSE (and consisting of $N=466$ companies) for the sub-period from 
2005-01-03 to 2006-03-09 (covering 309 trading days), which ranges distinctly before the worldwide financial crash. 
The companies are marked by small circles, except SALZGITTER AG -- Stahl und Technologie (SZG, marked 
by larger circle). 
We show below that this company plays a central role in the MST shown in Figure \ref{figure:20080812_asien}. 
If the link between two companies is in dark grey, the cross-correlation between them is large, while 
the distance between them is short (cf. the corresponding scale incorporated here). However, the geometric distances 
between companies, shown in this Figure by the lengths of straight line segments, are arbitrary, otherwise the tree 
would be much less readable.}
\label{figure:20060309_asien}
\end{center}
\end{figure}

The initial state of a graph representing the FSE is shown in Figure \ref{figure:20060309_asien}. In the following, 
we reveal the hierarchical structure of this graph and a corresponding scale-free MST. 

\begin{figure}
\begin{center}
\bigskip
\includegraphics[width=120mm,angle=0,clip]{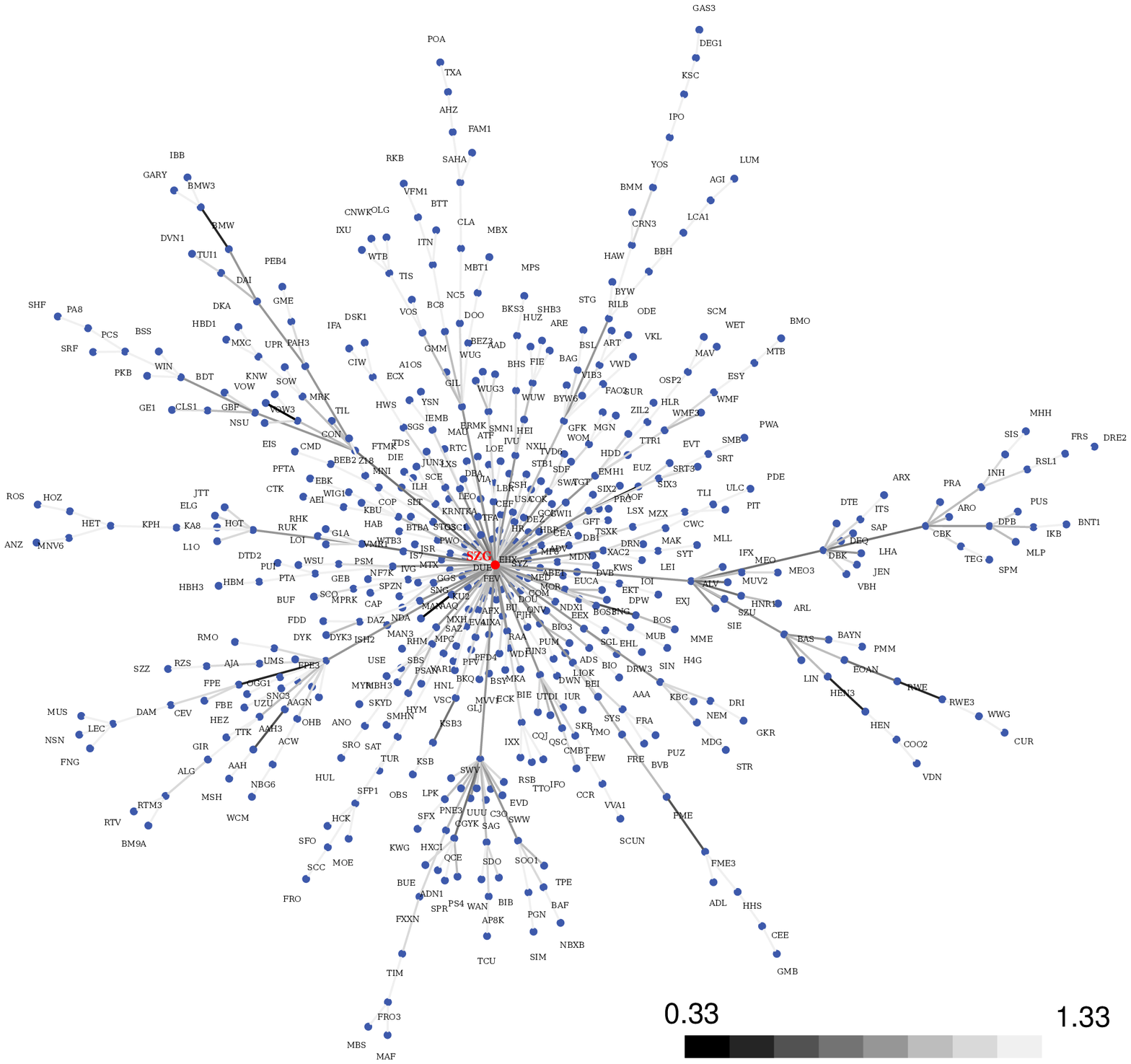} 
\caption{The superstar-like (or superhub) MST for the FSE (consisting of $N=479$ companies) observed for the sub-period 
from 2006-04-20 to 2007-10-31 (covering 400 trading days), which contains the worldwide financial crash. In this period the 
SALTZGITTER AG -- Stahl und Technologie company became a dominated hub (or superhub becoming also a giant component 
of the MST network), i.e. the central company of this stock market.}
\label{figure:20080812_asien}
\end{center}
\end{figure}

We focus our attention on the SALZGITTER AG -- Stahl und Technologie (SZG) company, which is a marginal company for the 
most of the period of time considered. However, it becomes a central company for the MST presented in Figure 
\ref{figure:20080812_asien}. This means that it is a central company only for the sub-period from 2006-04-20 to 
2007-10-31, which contains the worldwide financial crash. 
In other words, in this sub-period of time, the SALZGITTER AG -- Stahl und Technologie company is represented by a
vertex, which has a much larger number of edges (or it is of a much larger degree) than any other vertex (or company).
This means that it becomes a dominant hub (or superhub), also becoming a giant component of the MST. The company 
SALZGITTER AG -- Stahl und Technologie plays a role on the FSE, analogous to the company CAPITAL Partners on the WSE 
\cite{SGKS}. Similar role of the central node plays the General Electric (GE) among 116 companies of the S\&P 500 index 
for time period from 1983 to 2000 \cite{OCKK1,OCKK2}.

\begin{figure}
\begin{center}
\bigskip
\includegraphics[width=120mm,angle=0,clip]{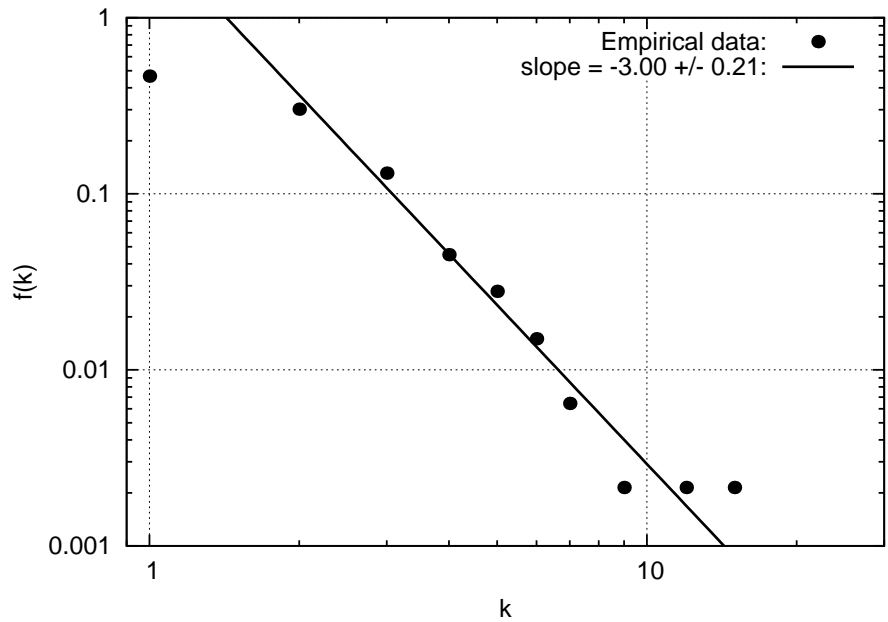} 
\caption{The power law distribution $f(k)$ vs. $k$ (where $k$ is the vertex degree) for the hierarchical scale-free 
MST network shown in Figure \ref{figure:20060309_asien}. Notably, we obtained here (i.e. for the period from 2005-01-03 
to 2006-03-09) the slope equal (to a good approximation) to that for the Barab{\'a}si--Albert complex network (which
equals 3.0). The number of companies taken into account in this period equals $N=466$. This plot is based on the window 
width $T=309\; td$. The results obtained for $T=350$ and $400$ are very similar, i.e. they are rather slowly-varying 
functions of $T$.}
\label{figure:spok_nie_spok}
\end{center}
\end{figure}

As described above, the transition between two structurally and topologically different states of the stock exchange is 
realized. We observe the transition from a scale-free MST (consisting of a hierarchy of local stars or hubs) to a 
superstar-like MST (or a superhub) decorated by a scale-free hierarchy of trees, that is, to the scale-free MST 
decorated by a temporal dragon king. 
The equivalent terms 'superextreme event' and 'dragon king' stress that \cite{DSDS}: (i) we are dealing with an 
exceptional event which is completely different in comparison with the usual events; (ii) this event is significant, 
being distinctly outside the power law. For instance, in paper \cite{WGKS} the sustained and impetuous dragon kings 
were defined and discussed.
Indeed, the MST shown in Figure \ref{figure:20060309_asien} leads to the distribution of vertex degrees in the form 
of a power law (see plot in Figure \ref{figure:spok_nie_spok}). 
\begin{figure}
\begin{center}
\bigskip
\includegraphics[width=120mm,angle=0,clip]{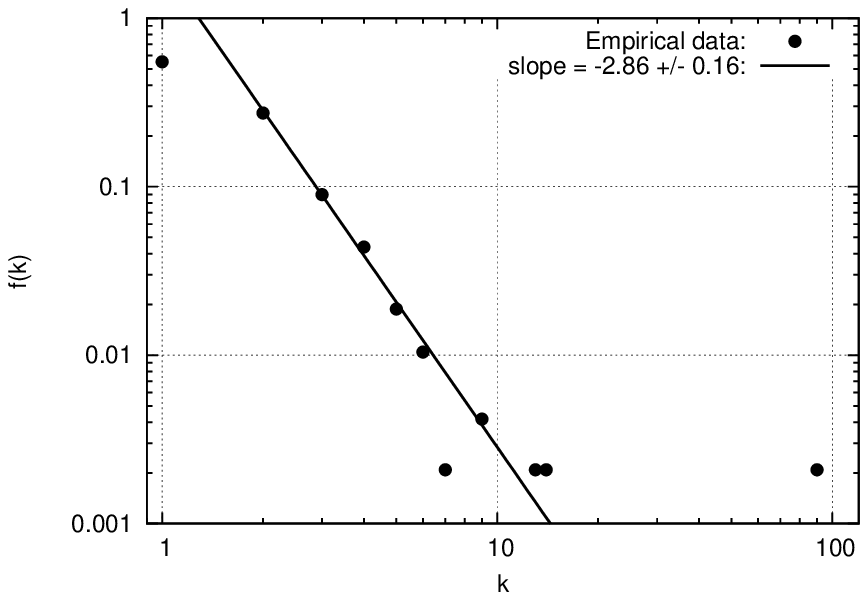} 
\caption{The power law discrete distribution $f(k)$ vs. $k$ (where $k$ is the vertex degree) for the superstar-like 
MST decorated by the hierarchy of scale-free trees shown in Figure \ref{figure:20080812_asien}. One can observe that 
there is a single vertex, which has degree $k=90$. This richest vertex represents the SALTZGITTER AG -- Stahl und 
Technologie company, which seems to form an extreme event -- the so-called dragon king \cite{DSDS,WGKS,AJK,MaSo}, in the period from 
2006-04-20 to 2007-10-31, being a giant component of the hierarchical scale-free MST network \cite{DGM}. The number 
of companies taken into account in this period equals 479. This plot is based on the scanning window of width $T=400\; td$. 
The results obtained for $T=350$ and $450$ are very similar, i.e. they are a slowly-varying function of $T$.}
\label{figure:spoko_nie_spoko}
\end{center}
\end{figure}
This degree distribution, $f(k)$, is described by a power law with an exponent $3.0\pm 0.21$ which is close, to a good approximation, to that of the Barab{\'a}si--Albert (BA) scale-free complex network (with their natural rule of preferential linking of new vertices) \cite{AB,DGM}. This exponent is distinctly larger than the corresponding one of the hierarchical MST on the WSE, which equals $1.97\mp 0.13$ \cite{SGKS}; coincidentally, this exponent value also characterises, e.g., the complex network of e-mails \cite{EMB}. Hence, the WSE appears to be more risky for stock market investments than the FSE, also for other time intervals (see below for details).

The power laws shown as well in Figs. \ref{figure:spok_nie_spok} and \ref{figure:spoko_nie_spoko} as below for Fig. \ref{figure:214_New} were found by the least square method, resultingin in a lower standard deviation than that obtained with the maximal likelihood fit. Furthermore, among several different fits using the least square method, the fit was chosen which again gives the lowest standard deviation. Namely, the fit to integer points from interval $[2, 10]$ were chosen. Plots based on scanning windows of widths $T=350$ and $450$, containing the same companies, gave very similar results, i.e. the results are rather slowly-varying function of $T$. In this (naive) sense, the sufficiently robust power laws were found.  

In Figure \ref{figure:spoko_nie_spoko}, the degree of distribution $f(k)$ vs. vortex degree $k$ is shown in a log-log plot. This distribution well fits a power law with a slope equal to $-2.86\pm 0.16$. This slope apparently is also characteristic of, e.g., the complex network of blogs \cite{BSLPRKMB} or MST for the Forex, where USD was assumed as the basic currency \cite{GDK}. However, the slope of the corresponding superhub on the WSE is driven by the distinctly larger slope equal to $-2.33\mp 0.17$; the latter slope is also characteristic for the complex network of actors (where also superstars are present) \cite{WS,ASBS} as well as for the Forex MST network where other than USD currencies were assumed as the basic ones (cf. Table 1 in Ref. \cite{GDK}).

The plot in Figure \ref{figure:spoko_nie_spoko} proves that the tree presented in Figure \ref{figure:20080812_asien} can be considered to be a hierarchical scale-free MST decorated by a dragon king. We hypothesize, that the appearance of such a dragon king is a signature of a stock market crash imminent within a few months (cf. ref. \cite{SGKS}). Obviously, this is a far going hypothesis which requires a systematic study.

Finding a proper local dynamical analytical description (perhaps nonlinear) for our network would be a formidable challenge. This can be appreciated by observing that the single vertex (representing the SALTZGITTER AG -- Stahl und Technologie company) is located far from the straight line (in the log-log plot) and can be considered to be a temporally outstanding, extreme event or a dragon king \cite{DSDS,WGKS,AJK,MaSo}, which condenses most of the edges (or links). Hence, the probability $f(k_{max})=1/479=0.0021$, where $k_{max}=90$ is the degree of the dragon king (which corresponds with the maximal degree here).

\begin{figure}
\begin{center}
\bigskip
\includegraphics[width=120mm,angle=0,clip]{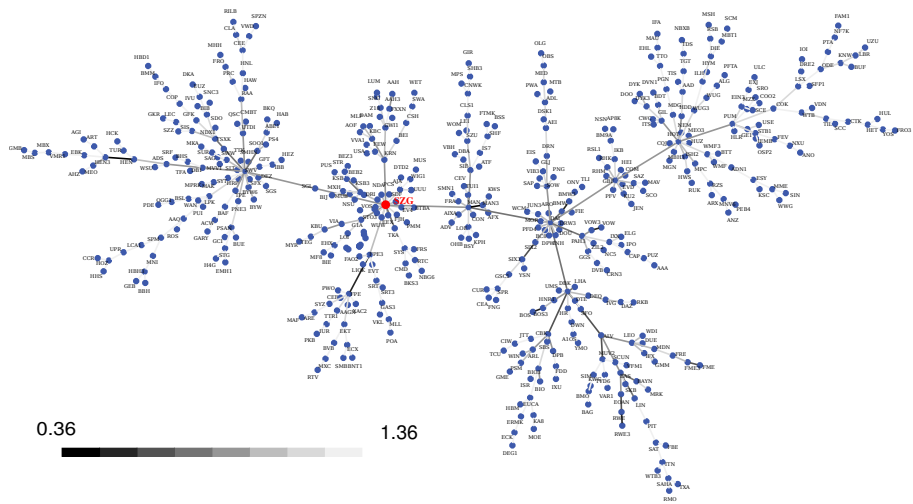} 
\caption{The hierarchical graph of the MST network decorated by several local star-like trees for the FSE for the period of time from 2007-06-01 to 2008-08-12 (covering 313 trading days). The SALTZGITTER AG -- Stahl und Technologie company is no longer a central hub. Where the link between two companies is in dark grey, the cross-correlation between them is greater, while the distance between them is shorter. However, the geometric distances between companies, reflected in the Figure by the length of the straight lines, are arbitrary, otherwise the tree would be much less readable.}
\label{figure:20080701-20110228_New}
\end{center}
\end{figure}

\begin{figure}
\begin{center}
\bigskip
\includegraphics[width=120mm,angle=0,clip]{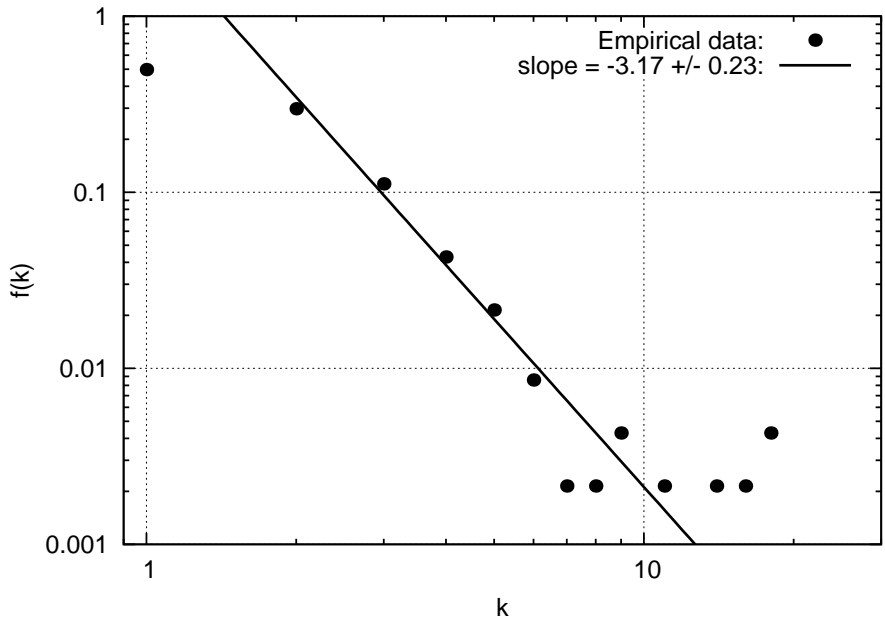}
\caption{The power law distribution $f(k)$ vs. $k$ (where $k$ is the vertex degree) for the hierarchical scale-free MST network shown in Figure \ref{figure:20080701-20110228_New}. This distribution was obtained for the period from 2007-06-01 to 2008-08-12. Three points (associated with four different companies) appeared above the power law. This means that four large hubs appeared instead of a single superhub. The number of companies taken into account in this period equals 479. This plot is based on the window width $T=313\; td$. The results  obtained for $T=350$ and $400$ are very similar, i.e. they are slowly varying with $T$.}
\label{figure:214_New}
\end{center}
\end{figure}

\begin{figure}
\begin{center}
\bigskip
\includegraphics[width=90mm,angle=270,clip]{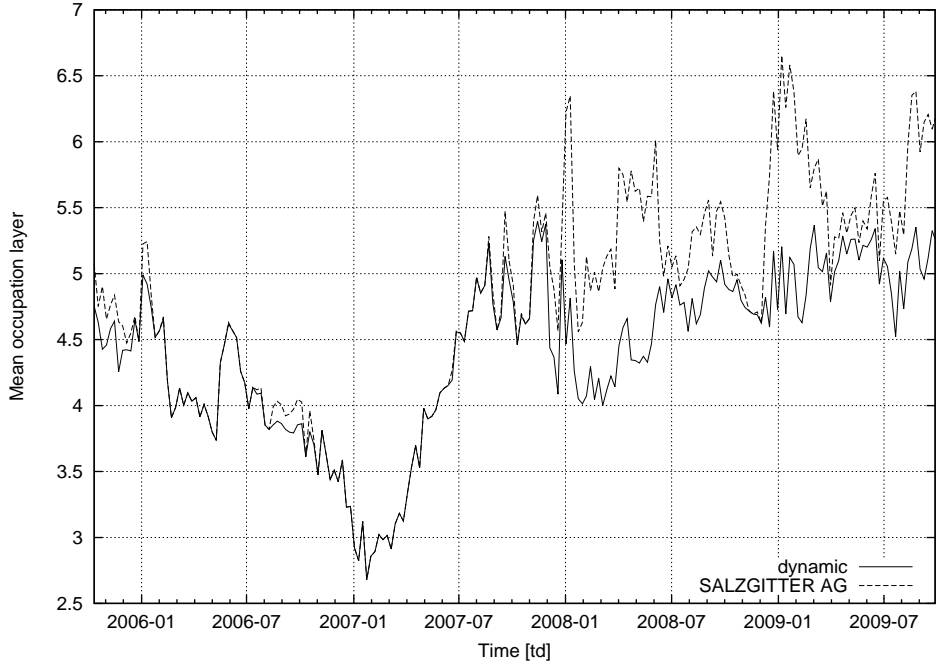} 
\caption{Mean occupation layer of MST vs. time (counted in trading days (td)). The result (marked by the solid curve) is based on such central temporal hubs, which have currently the largest degree. The central temporal hub means that this hub can be replaced from time to time by another central hub, during the stock market evolution. For comparison, the mean occupation layer was based on the SALZGITTER AG -- Stahl und Technologie company, assumed to be the central hub all the time (the dashed curve). A well defined absolute minimum, common for both curves, occurs at January 29, 2007. The region of coincidence of both plots extends from October 2005 to October 2009 indeed in the vicinity of the absolute minimum at January 29, 2007. Hence, we can conclude that the SALZGITTER AG -- Stahl und Technologie company is the central hub in this sub-period. The plot is obtained using the window width $T=400\; td$ and the time step of the scanning procedure equals $\delta t =5\; td$. The results obtained for $T=300$ and $350$ are very similar, i.e. they are a slowly-varying function of T.}
\label{figure:mean_layer}
\end{center}
\end{figure}

The sub-period of time containing the financial market crash was divided into two slightly overlapping intervals: the first one (already considered) from 2006-04-20 to 2007-10-31 and the second one from 2007-06-01 to 2008-08-12. Although both time intervals contain the worldwide financial crash, only the first MST is decorated by the dragon king, while for the second time interval MST is decorated by several intermediate hubs, somehow located between black swans and the dragon king (cf. Figures \ref{figure:20080701-20110228_New} and \ref{figure:214_New}). Hence, we can speculate that the dragon king appeared before 2007-06-01, which means it may have played the role of a crash precursor. (Further considerations concerning the dragon king localisation are indicated in the discussion of the plot in Figure \ref{figure:mean_layer}.)

\begin{figure}
\begin{center}
\bigskip
\includegraphics[width=120mm,angle=0,clip]{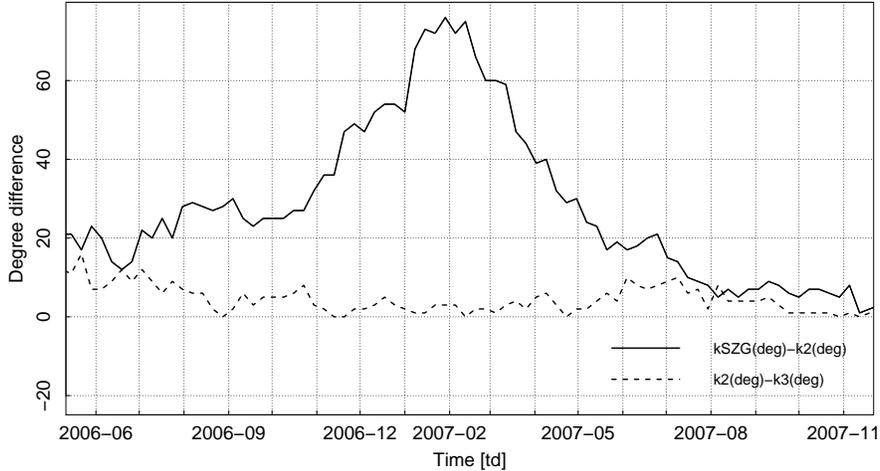} 
\caption{Empirical time-dependent increments $k_{SZG}(t)-k_2(t)$ (solid curve) and $k_2(t)-k_3(t)$ (dashed curve). Sudden increase of the former increment forming a steep peak in the vicinity of January 29, 2007 is well visible.}
\label{figure:degree_diff}
\end{center}
\end{figure}

To emphasize the increasing leading role of the SALTZGITTER AG -- Stahl und Technologie company we compare in Fig. \ref{figure:degree_diff} two different time-dependent increments: $k_{SZG}(t)-k_2(t)$ and $k_2(t)-k_3(t)$, where $k_{SZG}(t),\, k_2(t)$ and $k_3(t)$ are temporal degrees of the SALTZGITTER AG -- Stahl und Technologie company, vice-leader company and the third one in this rank. The accelerated increase of $k_{SZG}(t)-k_2(t)$ vs. time (the left-hand side of the peak plotted by the solid curve) observed, can in our case be considered as a signature of a dragon king, which appeared in the space of vertex degrees. Indeed, this results in a systematic (up to some relatively small fluctuations) expantion of SZG dynamics, ending in a superhub (or dragon king) star-like structure. This superhub is indeed a macro-transient structure, as it survives for only about half a year and decays afterwards (see the right hand side of the peak). Then MST becomes a modular structure consisting of few clusters (or sufficiently large star-like trees but not superhubs).
The dragon king dynamic equation, its solution and consequences are already considered in our subsequent work \cite{WSGKS}. In this context we also study there different time-dependent quantities of centrality and peripherality \cite{LCF,PB,MPA,PMA}.

It is interesting that several new hubs appeared for the period of time from 2007-06-01 to 2008-08-12, while a single superhub (superstar) disappeared, becoming a usual hub. This means that the structure and topology of the complex network significantly varies during its evolution through the market crash. This is well confirmed by the plot in Figure \ref{figure:214_New}, where several points representing large hubs (but not superhubs) are located outside the power law. Apparently, this power law is described by the slope equal $-3.17 \mp 0.23$. Hence, the consideration of this network as somehow equivalent to the Barab{\'a}si--Albert complex network is rather doubtful. 

Furthermore, the above given considerations are well confirmed by the plot shown in Figure \ref{figure:mean_layer}, where the clearly visible absolute minimum of the mean occupation layer is located at January 29, 2007 for the SALZGITTER AG -- Stahl und Technologie company, assumed to be the central hub.  

More precisely, the temporal (time-dependent) mean occupation layer of the MST, $mol[t; v_0(t)]$, represents, as usual \cite{OCKK1,OCKK2,OCKKK}, the mean number of subsequent edges connecting a given vertex of the MST, $v_j,\, j=1,\ldots , N$, with the temporal central vertex, $v_0(t)$, currently having the largest degree (and always counted as a null vertex). Hence,
\begin{eqnarray}
mol[t; v_0(t)]=\frac{1}{N}\, \sum_{j=1}^{N}lev[v_j(t), v_0(t)],
\label{rown:Level}
\end{eqnarray}
where $N$ is here the total number of vertices, less the temporal central one, and $lev[v_j(t), v_0(t)]$ is the current level of vertex $v_j(t),\; j=1,\ldots , N$, relative to $v_0(t)$. In other words, $lev[v_j(t), v_0(t)]$ is the current number of the MST edges linking (directly or indirectly) vertex $v_j(t)$ with the central one $v_0(t)$.\footnote{The remaining vertices assumed to be the central ones give at the same time shallower temporal minima.} This is a dynamic approach, as the central vertex may vary from one company to another one during the stock market evolution. Apparently, for a pure superstar-like MST, Equation (\ref{rown:Level}) gives $l[t; v_0(t)]=1$, as $lev[v_j(t), v_0(t)]=1,\, j=1,\ldots , N$. However, for the actual real-life situation shown in Figure \ref{figure:20080812_asien}, there also exist several subtrees (or local star-like trees), apart from the superstar-like tree, placed relatively far from the central vertex. Hence, it is not surprising, that the resulting time-dependent mean occupation layer exceeds $2.5$ (cf. Figure \ref{figure:mean_layer}) being still, however, sufficiently small. The mean occupation layer can be considered as the configurational weight of a macrostate of the complex network. This quantity is more sensitive to the complex network centrality than other quantities (e.g. the mean length used in our earlier work \cite{SGKS}) and its logarithm could be considered as a topological complex network entropy.

In Figure \ref{figure:mean_layer}, two essentially different types of predictions are shown. For the first type of prediction (dashed curve) the SALZGITTER AG -- Stahl und Technologie company was assumed to be the central hub all the time. Such an approach could be referred to as a static one. The second type of  prediction (solid curve) was obtained using the dynamic approach. Apparently, within the short sub-period ranging from 2006-11-01 to 2007-09-01 both approaches result in coinciding predictions, which coincidented. That is, the SALZGITTER AG -- Stahl und Technologie company is certainly the central hub in this sub-period (which is the main segment of the longer sub-period from 2006-04-20 to 2007-10-31). Indeed, the super-star like tree survived only in this segment. This is the most significant result of our work, indicating the existence of the most compact structure (mainly the super-star like one) at the beginning of 2007, but not at other times.  

We can consider the time-dependent mean occupation layer to be a time-dependent disorder parameter. For instance, this parameter indicates that the MST shown in Figure \ref{figure:20080812_asien} is less disordered than those shown in Figures \ref{figure:20060309_asien} and 
\ref{figure:20080701-20110228_New}, as expected.

\begin{figure}
\begin{center}
\bigskip
\includegraphics[width=130mm,angle=0,clip]{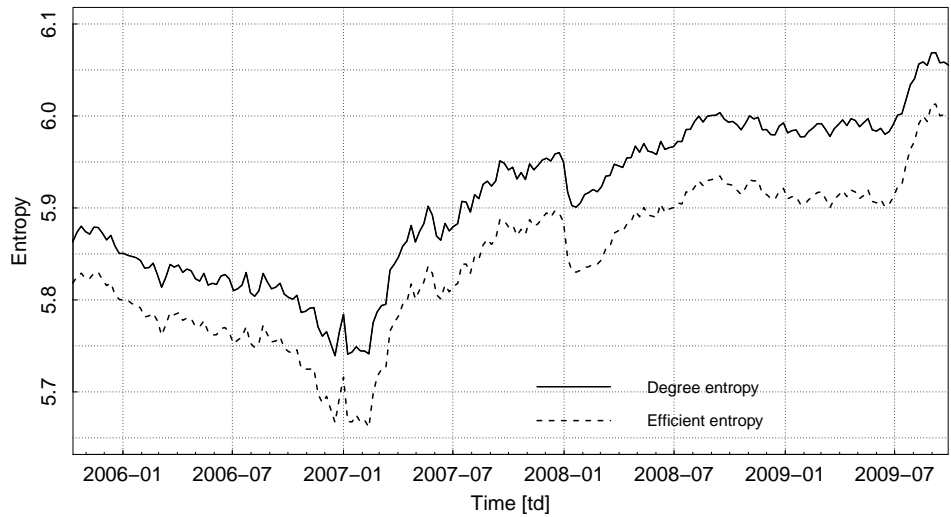} 
\caption{Two empirical entropies for MST of FSE vs. time (counted in trading days (td)). Stricking similarity of their shapes and MOLs can be observed, particularly in the vicinity of the absolute minimums, which all are located in January 29, 2007 (see Fig. \ref{figure:mean_layer} for details).}
\label{figure:entropy}
\end{center}
\end{figure}

To make confirmation of the solid network variation during its evolution, which relates to nonequilibrium statistical thermodynamics and information theory, we introduced two different time-dependent entropies. The first one, which we call degree entropy, $S_{deg}(t)$, is based on the empirical time-dependent degree distribution, $f(k,t)$. This entropy takes the form,
\begin{eqnarray}
S_{deg}(t)=-\sum_{k=1}^{k_{max}(t)}f(k,t)\ln f(k,t),
\label{rown:SDt}
\end{eqnarray}
where $k_{max}(t)$ is the maximal vertex degree at time $t$. The second entropy, which we call the efficient one, $S_{eff}(t)$, is based on inversed lengths of edges (or inversed distances between directly connected verticies\footnote{This form of entropy is inspired by complex network efficiency introduced in Ref. \cite{LM} in the context of small-world networks.}). It is defined as follows, 
\begin{eqnarray}
S_{eff}(t)=-\sum_{i=1}^n P(i,t)\ln P(i,t),
\label{rown:SMt}
\end{eqnarray}
where distribution of direct distances $P(i,t)=\sum_{j=1}^{k_i(t)}d^{-1}(i,j;t)/Norm(t)$, here $k_i(t)$ is a temporal degree of vertex $i$ and normalisation factor $Norm(t)=\sum_{i=1}^n \sum_{j=1}^{k_i(t)}d^{-1}(i,j;t)$. In Fig. \ref{figure:entropy} both empirical entropies were plotted vs. time. Apparently, the shape of both curves resemble those of MOLs shown in Fig. \ref{figure:mean_layer}. It is amazing how similar are both entropy curves. This unusual rebustness of the curve's shapes is likely caused by existence of the superhub. Furthermore, the absolute minimum of both entropies is (to good approximation) located at the same time as for MOLs. Hence, we suppose that observed features deeply relate structural and topologial properties of the network with their statistical thermodynamic and informational counterparts.

\begin{figure}
\begin{center}
\bigskip
\includegraphics[width=130mm,angle=0,clip]{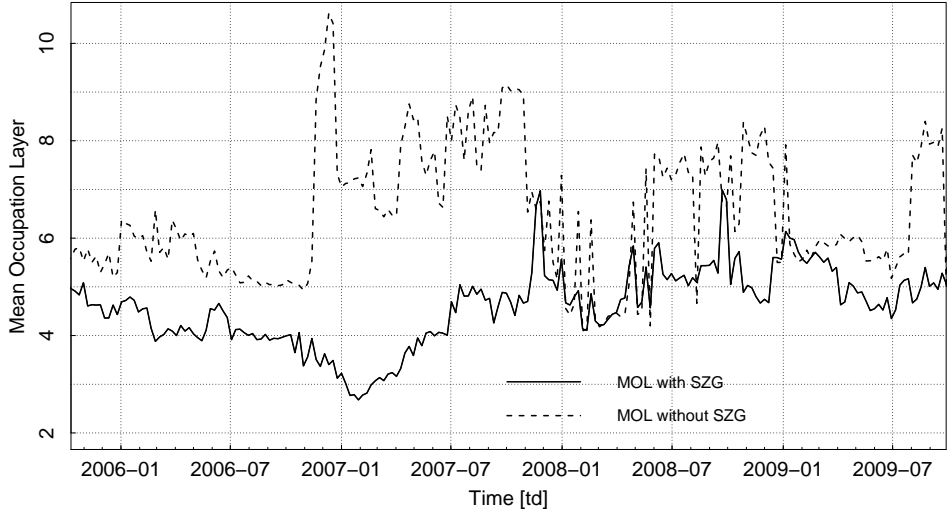} 
\caption{Two essentially diiferent MOLs vs. time (counted in trading days (td)). Stricking discrepancy between them is well seen in particular, in the vicinity of the absolute minimum of the MOL (solid curve) containing SZG company (i.e. in the vicinity of January 29, 2007).}
\label{figure:2_mols}
\end{center}
\end{figure}

To be sure that SZG company is responsible for the existence of the absolute minimum of the MOL at January 29, 2007 (shown in Fig. \ref{figure:mean_layer}), 
we compare in Fig. \ref{figure:2_mols} to essentially different MOLs: (i) the above mentioned MOL and the second one (ii) calculated for the modified FSE, where the SALZGITTER AG -- Stahl und Technologie was absent. Apparently, the behaviour of both curves is drastically different and no absolute minimum (as defined above) is present for the second MOL.  

\section{Concluding remarks}

In this work, we have studied the empirical evolving correlated network associated with the FSE - a stock exchange of medium capitalisation. Our results reveal a surprising fact that a company of a medium size (herein SALZGITTER AG -- Stahl und Technologie) becomes a dominant hub -- a superhub -- in the critical, metastable regime (i.e. in the second period considered, cf. Figure \ref{figure:20080812_asien} for details). This is because correlations between this company and other more significant ones likely become weaker in the vicinity of the crash regime. This results in the relative increase of the correlations with less important companies, such as, for instance, the SALZGITTER AG -- Stahl und Technologie -- which also remains relatively unaffected within the entire period of the market evolution considered here.

Our work provides an empirical evidence of a dynamic structural and topological first order phase transition (or discontinuous one) in the time range dominated by a stock market crash (that is, from 2006-04-20 to 2007-10-31). Before and after this range, the superhub (or unstable state of FSE) diappears and we respectively observe either a pure hierarchical, scale-free MST or a hierarchical MST decorated by several hubs. Hence, our results consistently confirm the existence of the following dynamic structural and topological phase transitions, which can be briefly summarized as folows: 
\begin{eqnarray}
& &\mbox{phase of scale-free MST - a (relatively) stable stock market state} \nonumber \\ 
&\Rightarrow &\mbox{phase of the superstar-like MST - a transient market state} \nonumber \\ 
&\Rightarrow &\mbox{phase of scale-free MST decorated by few local star-like trees} 
\nonumber \\
& &\mbox{- a (relatively) stable stock market state}.
\nonumber
\end{eqnarray}
We hypothesise that the first of these transitions can be considered to be a precursor of a financial crash as it appeared a few months before this crash. 

Another significant observation presented in this work appertains to the power laws plotted in Figures \ref{figure:spok_nie_spok}, \ref{figure:spoko_nie_spoko}, and \ref{figure:214_New}. Namely, the exponent of degree distribution presented in Figure \ref{figure:spoko_nie_spoko} is, in fact, distinctly smaller than $3$, which means that the variance of this vertex degree diverges. This result indicates that we are here dealing with criticality, which means that the scenario of our network evolution takes place within a scaling region \cite{DGM,DS0,BaPo,HH} containing critical phenomena.

Our results are to some degree complementary to those obtained earlier by Dro\.zd\.z, Kwapie\'n and Speth \cite{DrKw}. Their results focused on the slow (stable) component (or state). Namely, they constructed the MST network of 1000 highly capitalized American companies. The topology of this MST show its centralization around the most important relatively stable node, it being the General Electric. This was found both in the frame of binary and weighted MSTs.

In this context, the fact should be stressed that the discontinuous phase transition (i.e. the first order phase one) topologically preceds and develops into the continuous phase transition (i.e.  the second order one). This discontinuous phase transition goes over the unstable state involving, perhaps, a superheated state such as the superhub in our case. This cannot be considered as noise\footnote{Indeed, the case of the noise was discussed in this context in paper \cite{STZM}.} in the system, but rather should be considered as a result of the natural evolution of the system until the critical point is reached (cf. \cite{KwDr} and refs. therein, where the role of stable states (or slow components) on NYSE or NASDAQ was considered by using binary and weighted MSTs).

In this work we also studied (but did not visualize) the stock market evolution for the subsequent (the fourth) time interval from 2008-07-01 to 2011-02-28 and found the power law degree of distribution driven by exponent equals $2.82\pm 0.36$. In addition, the power law was decorated by a few points located far above this law. 
This likely indicates, that the system after leaving one critical regime, is approaching a new one. 

We suppose that the phenomenological theory of cooperative phenomena in networks proposed by Goltsev et al. \cite{GDM} might be a promising first attempt to investigate this kind of structural topological dynamics of criticality. An alternative view might consider the superhub phase to be a temporal condensate \cite{DGM}. Hence, we can briefly reformulate the above-mentioned phase transitions as representing the dynamic transition from the excited phase into the condensate and then the transition outside of the condensate to an excited phase again. 

Obviously, an analytical treatment of the dynamics of such a network phase transition, in particular the superstar-like tree formation, is a formidable challenge. We can conclude this work with the hope that the continued detailed study of phase transitions presented may define the basis for a better understanding of a stock market crash dynamics and a basis for theory development.

\section*{Acknowledgments}

We are grateful Rosario N. Mantegna and Tiziana Di Matteo for helpful comments and suggestions.

\end{document}